# Shaping Near Fields of Photonic Modes into Topological Textures

*Jonas Biechteler*[*,1,2]*, Michael Hirler*[*,1,2]*, Connor Heimig*[1,2]*, Martin Barkey*[1,2]*, Haoran Ren*[3]*,*

*Stefan A. Maier*[3,4]*, Enrico Baù*[‡,2] *and Andreas Tittl*[†,1,2]

[1]Institute of Photonics, Hamburg University of Technology, 21073 Hamburg, Germany.

[2]Chair in Hybrid Nanosystems, Nanoinstitute Munich, Department of Physics, Ludwig-Maximilians-Universität, 80539 Munich, Germany.

[3]School of Physics and Astronomy, Monash University, Clayton, Victoria 3800, Australia.

[4]Department of Physics, Imperial College London, London SW7 2AZ, United Kingdom.

*These authors contributed equally to this work.

[†]Email: andreas.tittl@tuhh.de

[‡]Email: enrico.bau@physik.uni-muenchen.de

**ABSTRACT**

Topological textures of light, such as optical vortices and skyrmions, are attractive for high-density information encoding and light-matter interactions. In the near field, topologies have primarily been realized using polaritonic platforms that inherently tie the platform to the material dispersion, limiting their footprint due to propagation losses and restricting their operating spectral region. Here, we introduce a route to shape near-field topologies through the geometrical engineering of a dielectric metasurface supporting a quasi-bound state in the continuum resonance, without the need for polaritonic surface waves. In-plane deformation of our metasurface bars conserves their resonance-driven near fields locally, so that the metasurface geometry and the polarization of the illuminating light dictate the generated topology. By shaping silicon unit cells into geometries with rotational symmetry, we create optical topologies with controllable topological charge. Using phase-resolved scattering-type near-field optical microscopy, we experimentally characterize high-purity resonance-locked topologies such as vortices in concentric rings and topological charge doublets in Archimedean spirals. In contrast to conventional topological textures, our purely photonic mode based field textures are stationary standing waves in the radial direction, without relying on surface-wave propagation or polaritonic dispersion, establishing a spectrally and spatially scalable platform for near-field topological photonics.

## INTRODUCTION

Optical metasurfaces are ultrathin, planar arrays of subwavelength structures designed to manipulate light down to the nanoscale.[1–3] Owing to extensive research in recent years, metasurfaces encompass a wide variety of geometries[4,5] and operating principles[6,7] that enable precise control over optical amplitude,[8–10] phase,[1,8,10–12] polarization,[13,14] and dispersion[15–17] across both near- and far-field regimes.

In particular, metasurfaces have entered the field of topology as a versatile platform for exploring topological phenomena. Owing to their ability to spatially vary phase profiles, metasurfaces can engineer far-field waves to carry topological features, as demonstrated for optical vortex beams carrying angular momentum (AM),[1,18,19] radially and azimuthally polarized vector vortex beams,[20] Stokes skyrmions[21,22] and arbitrary spin-to-orbital angular momentum state converters[23]. Non-local metasurfaces can also exploit the momentum-space polarization vortex that surrounds a symmetry-protected bound state in the continuum (BIC) to imprint a geometric phase onto the outgoing beam.[19,24,25] Nevertheless, far-field topologies are fundamentally bounded by the diffraction limit.

In the near field, by contrast, evanescent fields can vary on scales far below the wavelength. Research has therefore expanded to near-field polaritonic topological textures, such as localized optical vortices[26–30], polaritonic skyrmions[31–34] and merons[35,36]. These polaritonic structures enable strong spatial confinement and feature sizes well below the diffraction limit.[32] Both characteristics have motivated proposals for high-density information encoding,[37,38] optical information processing,[39] and the emulation of condensed-matter phenomena[40].

However, common approaches relying on surface plasmons or phonon polaritons require interfaces between a negative-permittivity material and a dielectric.[41,42] As a consequence, the platforms inherit the constraints of the underlying materials and the operating wavelength is confined to a narrow spectral window where the material's permittivity is negative, preventing broadband spectral scalability and heavily restricting the choice of material for a given target wavelength.[43,44] Furthermore, the propagating nature of polaritons inherently limits the spatial extent of the texture and the achievable device footprint due to their propagation losses.[35,45] Recent work has shown that non-local photonic modes can be leveraged to generate polaritonic topologies, including tunable skyrmion lattices and AM-carrying vortices.[30,34] Nonetheless, the resulting field is still carried by the polariton and therefore remains subject to these constraints. Geometry, meanwhile, offers a degree of freedom that is independent of the material response: shaping the device layout is an established route to, for example, create and guide spoof surface plasmons via the establishment of a geometry-induced Drude plasma dispersion[46,47] or to extend metasurface functionality,[48–51] but its use to shape a purely photonic resonant near-field topology remains largely unexplored.

Here, we conceptually develop and experimentally demonstrate a versatile design approach based on the in-plane deformation of the resonating elements constituting the metasurface, generating topological textures directly in their near field. Unlike polaritonic topologies, our approach creates near-field textures that are generated purely by a photonic mode and set by the metasurface geometry, making them independent of polaritonic dispersion and the material constraints it imposes. Figure 1a shows the schematic of a dielectric one-dimensional (1D) displacement grating, a well-established geometry[52,53] in which parallel bars form a periodic array. Without displacement, this array supports a guided mode that lies below the light line and is therefore decoupled from free space. Displacing every other bar doubles the period and folds this mode into the radiative continuum, turning it into a Brillouin zone folding quasi-BIC (qBIC)[54] that achieves strong field confinement.[55,56]

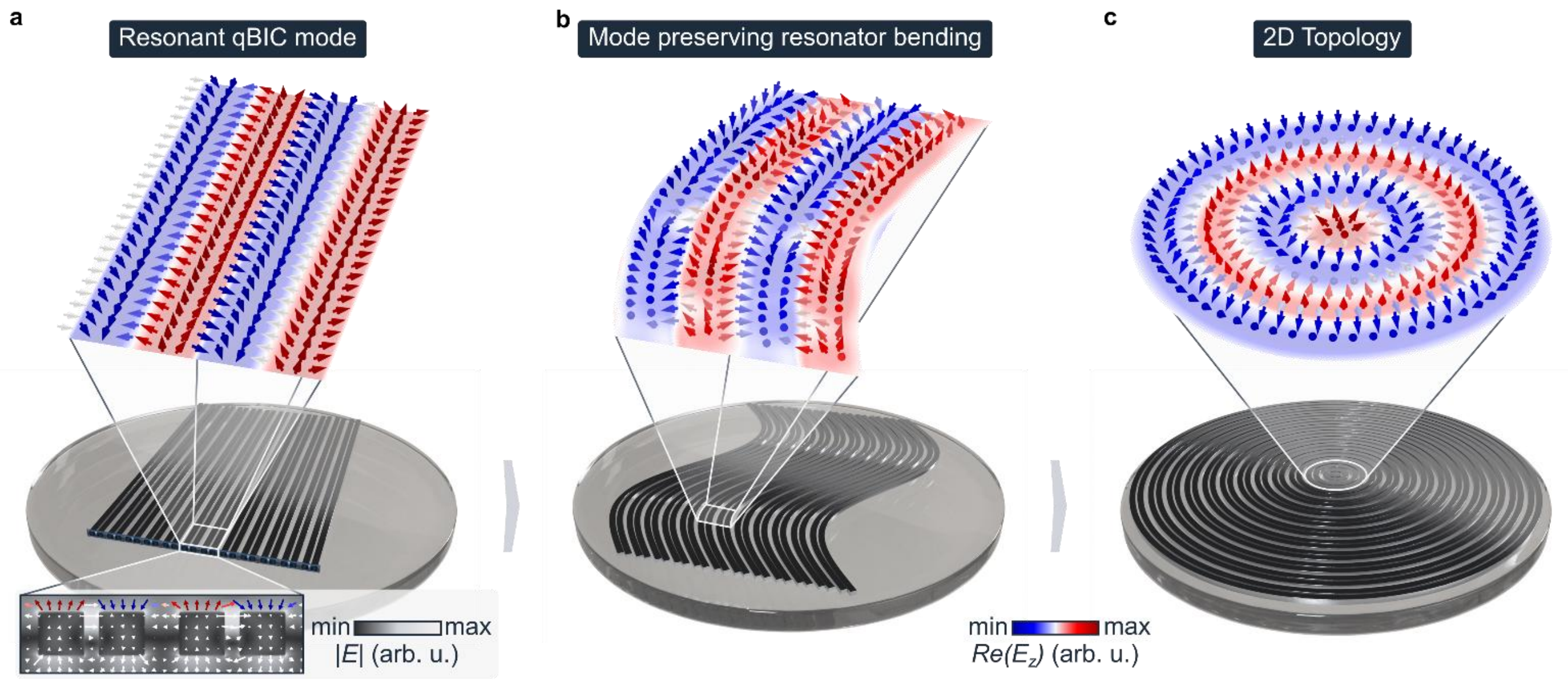

**Figure 1**. 1D metasurface bending unlocks topologies. (a) 1D displacement grating supporting a non-local, resonant photonic qBIC mode. The inset (bottom) shows the simulated electric near-field profile, while the field texture (top) is depicted in the plane 5 nm above the resonator bars. (b) Bending the resonator bars in-plane (here in a sinusoidal shape) locally preserves the field resonant mode. (c) Deliberate bending enables the generation of well-defined topological textures such as target skyrmions, here via concentric rings under radially polarized excitation.

In the plane directly above the resonators, it provides a non-local and coherent field texture that is continuously rotating orthogonally to the bar direction, well suited for the realization of topological textures. We evaluate the field texture 5 nm above the resonator surface, corresponding to the experimentally accessible region (see Methods).

Bending the bars in the metasurface plane with a radius of curvature much larger than the array period (Figure 1b) preserves the qBIC resonance locally and causes the near-field texture to conform to the bar trajectory. The curvature of the bars thereby becomes a design parameter that maps directly onto the phase distribution of the resonant near field we set out to shape. We exploit this principle to generate topological near-field textures by design. Figure 1c shows the example of a metasurface of concentric rings, which can support a Néel-type optical target skyrmion under radially polarized excitation (see Results and Discussion).[57]

We investigate the qBIC mode formation in concentric ring metasurfaces and experimentally demonstrate vortices with AM, characterized via transmission-mode scattering-type scanning near-field optical microscopy (s-SNOM). Generalizing the bending principle to create Archimedean spiral metasurfaces allows encoding of arbitrary geometry-controlled AM doublets into the near fields, which we demonstrate with up to 20 unit-cell arms. The approach thereby opens a route to compact near-field information encoding and on-chip topological photonics, by virtue of the platform's free scalability down to wavelengths in the ultraviolet (UV).

## RESULTS AND DISCUSSION

### Mode formation in concentric ring metasurfaces

Throughout this work, all metasurfaces share the same design principle: Curved bars with alternating gap dimensions that support qBIC resonances. Figure 2a illustrates this for a metasurface consisting of radially periodic concentric rings, made from silicon (Si) bars on a mid-infrared (IR) transparent calcium fluoride ($CaF_2$) substrate. The platform hosts two non-local modes, one with strong in-plane magnetic and another with a strong electric dipole contribution.[58] We classify them as 'out-of-plane' and 'in-plane' modes, owing to their pronounced electric field components in the respective direction (Figure 2b). For linearly polarized light, the incident electric field can be decomposed locally into a perpendicular component oriented across the bars and a parallel component oriented along them.

The out-of-plane mode couples predominantly to the perpendicular component, whereas the in-plane mode couples to the parallel one (Supplementary Note 1). Because circularly polarized (CP) light provides perpendicular and parallel components with equal magnitude at every azimuthal angle, the two modes are excited uniformly across the rotationally symmetric metasurface.

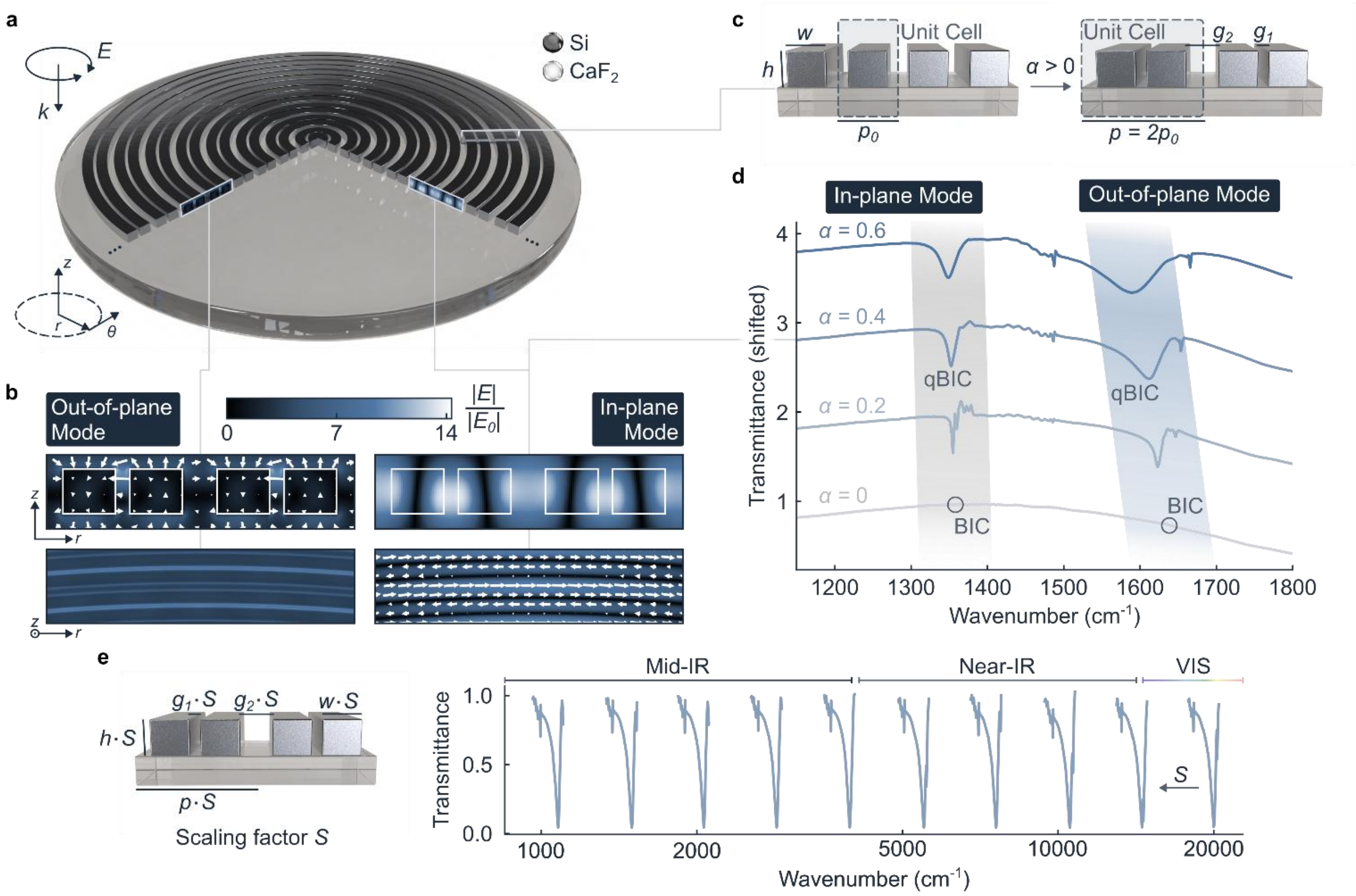

**Figure 2**. Concentric ring metasurface supporting displacement-mediated qBICs. (a) Artistic rendering of the metasurface with a cutout, excited with CP light. (b) Simulated electric near-field enhancement of the out-of-plane and in-plane modes for α = 0.4. (c) The metasurface can be locally approximated as a 1D superlattice featuring two bars per unit cell with different gaps $g_1 \neq g_2$ for asymmetry parameters $\alpha > 0$. (d) Simulated far-field transmission spectra for different asymmetries, illustrating the emergence and linewidth tuning of the qBIC resonances as a function of the asymmetry $\alpha$. (e) Resonance scaling via scaling parameter S (here from 0.08 to 1.5), shown for the out-of-plane mode from VIS to mid-IR wavenumbers.

For both modes, the field is efficiently concentrated within the gaps, making the platform a promising candidate for effective light-matter coupling.[58,59]

Beyond exhibiting intense near-field hotspots, the gap dimensions provide a key structural parameter for tailoring modal properties. The metasurface can be locally approximated as a 1D displacement grating with two bars per unit cell. Figure 2c shows a cross-sectional view of two adjacent unit cells. We label the gap between the two bars within a unit cell $g_1$, and between bars of neighboring unit cells $g_2$. All bars have a uniform height $h$ = 1230 nm, a width $w$ = 1300 nm and the unit-cell period is $p$ = 4000 nm. On this basis, we define the asymmetry parameter $\alpha = (g_2 - g_1)/(g_2 + g_1)$, which serves as a control handle for the radiative loss channel of the system. For $\alpha = 0$, the unit cell comprises a single bar and supports a guided mode residing outside the radiative continuum and unable to couple to free-space light due to total internal reflection. For $\alpha > 0$, the bar displacement doubles the period and halves the size of the first Brillouin zone in momentum space, folding the guided mode into the radiative continuum. Simultaneously, the displacement lifts the translational symmetry such that the Brillouin zone folding BIC can couple to radiative channels and turns into a qBIC.[54] In the far field, the in-plane and out-of-plane modes emerge as sharp resonances with finite but high quality ($Q$) factors at around 1350 and 1620 $cm^{-1}$ (Figure 2d). We design our metasurface to operate in the mid-IR, where we have hyperspectral far-field and phase-resolved near-field imaging capabilities. By fine tuning the gap ratio, the resonance linewidth can be deliberately tailored, satisfying the well-known relation $Q \propto \alpha^{-2}$ in the absence of intrinsic losses (Supplementary Note 2).[56] The spectral resonance position, in turn, can be controlled by multiplying all unit-cell dimensions with a scaling factor $S$. Using a dielectric like Si allows for broadband spectral tuning, where the tunability is bounded solely by fabrication constraints and the onset of intrinsic material losses at shorter wavelengths.[59] Figure 2e showcases the scalability of the out-of-plane mode, with resonances spanning from the visible (VIS) to the mid-IR spectrum.

Throughout the remainder of this work, we focus on the out-of-plane mode. In contrast to the in-plane mode, its electric field vector in the plane above the resonators covers all orientations in the plane perpendicular to the bars within one period, which can be used for the generation of near-field topological textures.

We exploit this near-field distribution to generate a Néel-type optical target skyrmion in simulation using radially polarized excitation, as previously shown in Figure 1c.[57] The radial polarization drives the out-of-plane mode with the same phase at every azimuthal angle, and the real-space field vector above the metasurface rotates coherently in radial direction. It points predominantly up- or downwards above one resonator bar, turns in-plane across the gap, and points into the opposite direction above the adjacent bar. To investigate the topological nature of our near-field texture, we evaluate the skyrmion number density $s$, which quantifies the spatial distribution of the near field's topological characteristics:

$$s = \frac{1}{4\pi} \hat{e} \cdot \left(\frac{\partial \hat{e}}{\partial x} \times \frac{\partial \hat{e}}{\partial y}\right), \tag{1}$$

where ê denotes the normalized electric field vector. Then we integrate it from the center of one resonator bar at $R_n$, to the center of the adjacent bar at $R_{n+1}$, to calculate the respective topological winding or skyrmion number $S_T$, which evaluates the amount of times the vector field wraps around the unit sphere:[31]

$$S_T = \int_{R_n}^{R_{n+1}} s \, dA. \tag{2}$$

We find rotationally symmetric skyrmion rings of width $p/2 \pm (g_2\text{-}g_1)/2$ with a charge of $S_T = \pm 1.00 \pm 0.025$, with their sign alternating from ring to ring (see Supplementary Note 3). Contrary to polaritonic skyrmions, which are carried by propagating surface waves, the spatial near-field texture of our skyrmions remains pinned to the underlying silicon resonators and forms a radial standing wave, since only its amplitude oscillates in time. Each skyrmion ring stays stationary and only reverses its polarity every half period over one optical cycle. Owing to the non-local nature of the underlying photonic mode, the skyrmion can be extended by an arbitrary number of rings.

**Generation of resonant near-field vortices**

For the experimental investigation of the proposed concentric ring metasurface design, we fabricate the structures from amorphous Si on $CaF_2$ following the procedure described in Methods, using previously simulated parameters ($\alpha = 0.4$, $p = 4000$ nm, 30 unit cells). The scanning electron microscopy (SEM) image in Figure 3a shows the central part of the metasurface, revealing the radially periodic and rotationally symmetric design with high structural uniformity. We first monitor the out-of-plane resonance using a hyperspectral imaging microscope that records transmittance spectra at every pixel under CP light (see Methods). Along the azimuthal direction, we integrate and average the spectra over 10° intervals and plot the resulting spectra against the azimuthal angle $\theta$ in Figure 3b. The out-of-plane mode shows an average $Q$ factor of ~50 and sufficient spectral stability with fluctuations comparable to the resonance's full width at half maximum. Figure 3c displays the measured transmittance spectra in a radial cut. Integrating a 20 µm wide domain along the radial axis in 5 µm steps across the entire metasurface reveals the out-of-plane mode behavior throughout the edge and center areas. Towards the inner rings, where the deformation of the Si bars becomes increasingly strong, the resonance contrast decreases, but the resonance persists.

For the extraction of the near-field textures generated by the ring metasurfaces, transmission-mode s-SNOM was employed,[60–62] as Figure 3d illustrates. The entire metasurface is illuminated with CP light at normal incidence at 1620 $cm^{-1}$ to excite the out-of-plane mode. Both the out-of-plane near-field phase $\varphi_z$ and amplitude $|E_z|$ are extracted simultaneously via a pseudo-heterodyne detection scheme[63] (see Methods). Figure 3e shows the resulting near-field maps of phase (top) and amplitude (bottom) for the three innermost ring pairs under left circularly polarized (LCP) excitation after Fourier filtering to improve visual clarity of the experimental data (Supplementary Note 4) following a previously developed approach.[64] The maps show a pattern periodic in radial direction with full π phase jumps from one ring to the adjacent one, while in azimuthal direction the phase sweeps through a full 2π over one rotation. On the disk in the center of the ring metasurface, a phase singularity forms. The near field thus forms an optical vortex, driven entirely by the qBIC resonance of the dielectric metasurface.

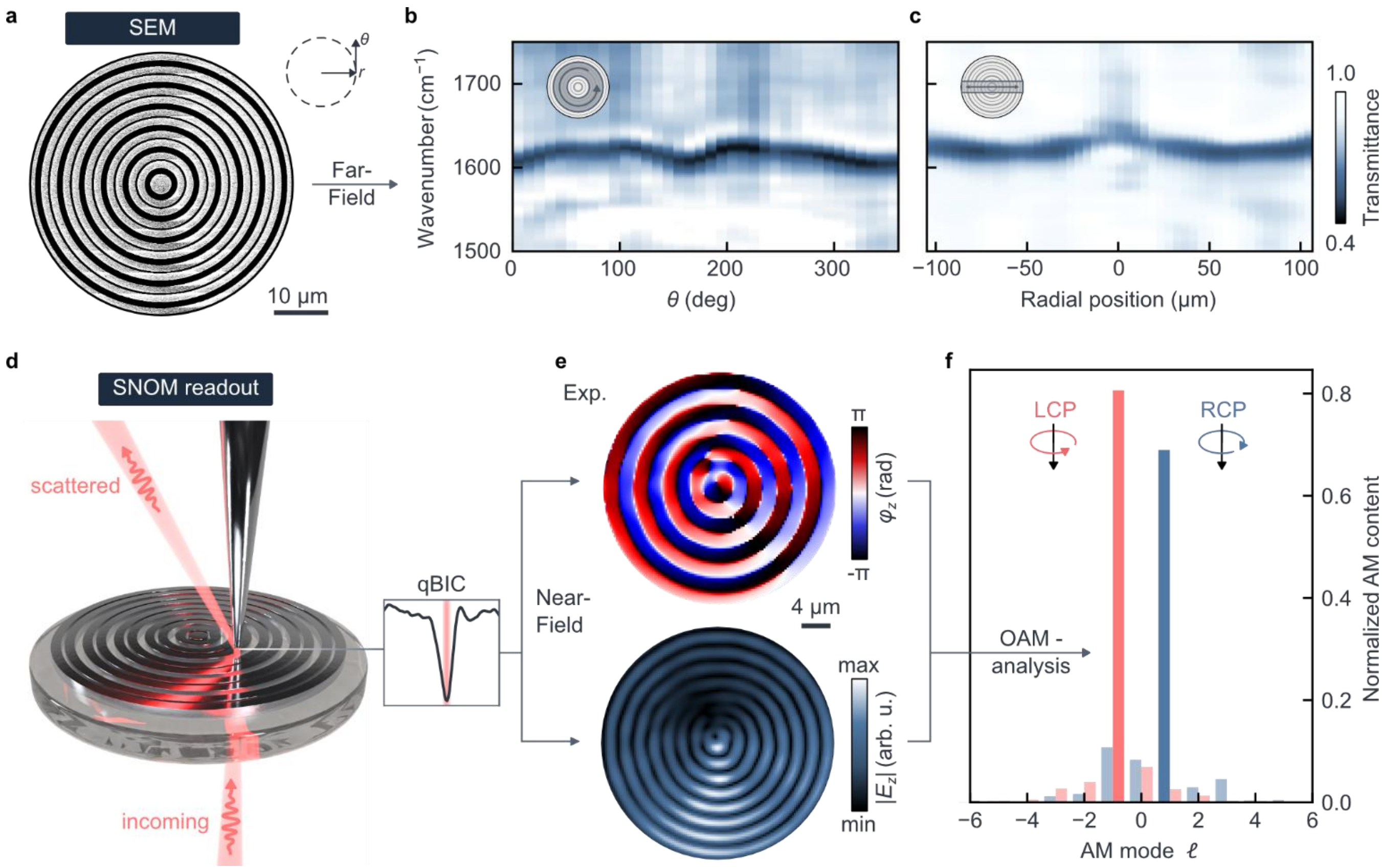

**Figure 3**. Far- and near-field measurements of concentric ring metasurfaces. (a) SEM image of a concentric ring metasurface. (b) Far-field spectra under rotation in azimuthal direction θ. (c) Far-field measurements along the radial axis through the center of the metasurface. (d) Artistic rendering of the s-SNOM measurement of the evanescent fields generated when the qBIC mode is resonantly excited. (e) Experimental near-field images of phase (top) and amplitude (bottom) using LCP light. (f) Normalized AM power spectra for different handednesses of incoming circular polarization.

The phase gradient is carried by the continuous curvature of the out-of-plane mode and does not propagate, as it is bound to the resonant structure below. To analyze the observed optical vortex, we apply an azimuthal Fourier decomposition to the measured fields (see Supplementary Note 5) and find that the near fields predominantly carry the topological charge $\ell$ that is solely imposed by the handedness $\sigma$ of the incoming light as $\sigma = \ell = \pm 1$, with -1 for LCP and +1 for right circularly polarized (RCP) light illumination and $\ell$ indicating the number of phase windings by $2\pi$ around the vortex center (Figure 3f).

The generation of this near-field vortex is tied to the excitation of the qBIC. Elimination of the asymmetry ($\alpha = 0$) leads to complete absence of the qBIC resonance and of the desired field texture, as shown in Supplementary Note 6.

## Geometric control of the near-field angular momentum

To achieve higher topological charges, we take our mode-preserving bending principle one step further to reshape the ring design into an Archimedean spiral. Here, a single pair of bent resonators functions as the spiral arm by winding out of the center of the metasurface with a radial offset of $p$ per full revolution (Figure 4a). An SEM image of a fabricated one-armed ($N = 1$) spiral metasurface is displayed in Figure 4b. The geometrical switch from the ring to the spiral design keeps the radial periodicity and thereby sustains the out-of-plane mode formation under CP illumination. Figure 4c shows the spectral stability of the qBIC resonance for a full azimuthal rotation along $\theta$. The spiral therefore supports the same non-local mode as the rings, since the resonance is set by the local unit cell while the topology is governed by how that cell is laid out globally. In contrast to previous approaches,[64] there is no need to match the spiral pitch to the polariton wavelength. Figure 4d and e show the measured near-field phase map (top) and amplitude (bottom) across the inner four full windings of the $N = 1$ spiral metasurface under LCP and RCP excitation, respectively. The phase pattern resembles the geometry of the underlying metasurface, forming a spiral. To investigate the phase evolution of the platform in time, we multiply the pattern with a phase factor $e^{i\varphi}$, with $\varphi$ looping from 0 to $2\pi$. This reveals a standing wave-like character of the field texture in radial direction. Here, the radial field profile stays stationary throughout one optical cycle, while its amplitude oscillates. This oscillation does not happen coherently at all angles but follows a rotating motion in azimuthal direction.

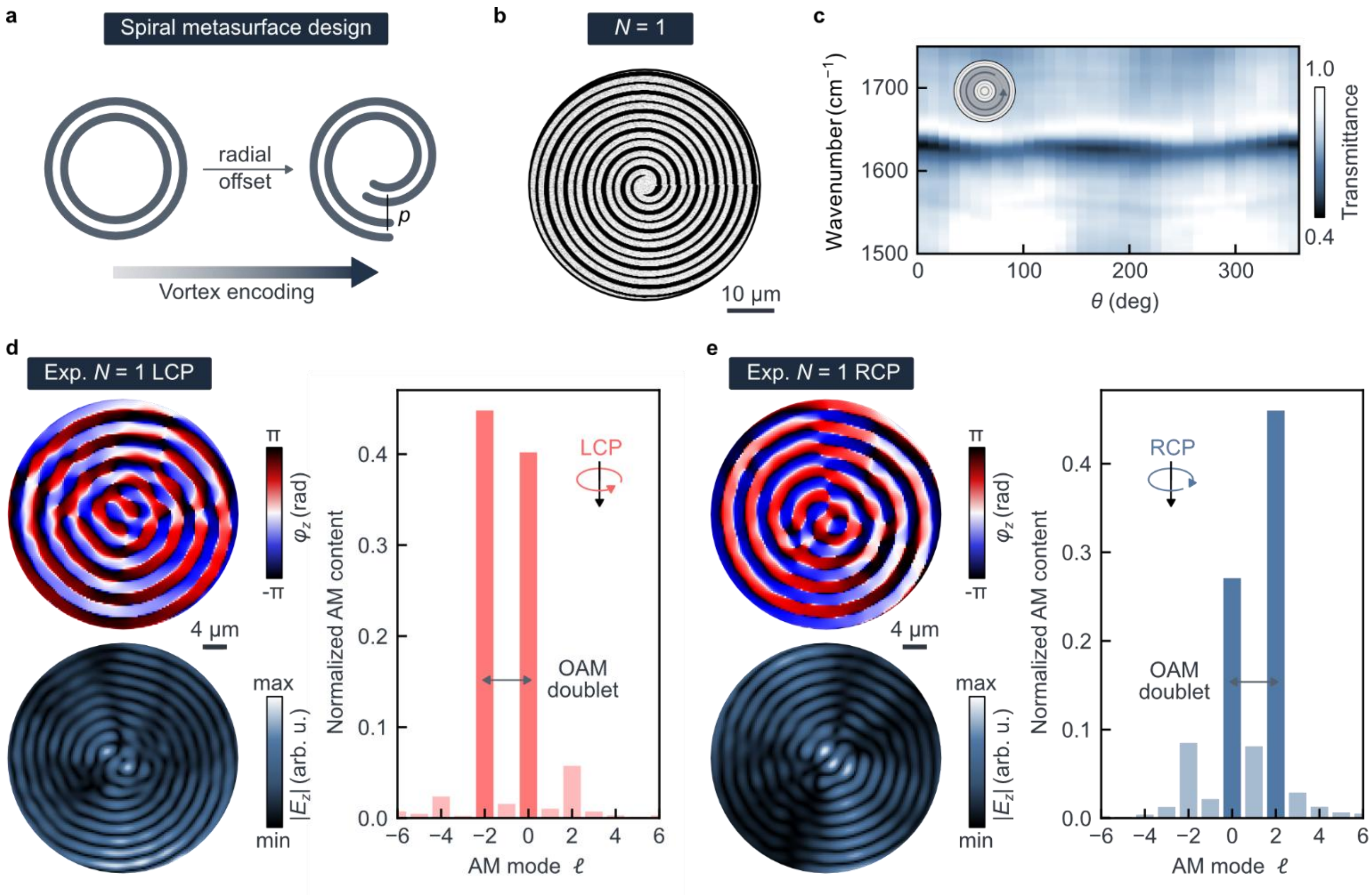

**Figure 4**. Far- and near-field measurements of an Archimedean spiral metasurface. (a) Schematic illustration of the transition from ring to Archimedean spiral metasurface for near-field vortex generation. (b) SEM image of the single-armed (N = 1) spiral metasurface. (c) Far-field spectra under rotation in azimuthal direction θ. (d, e) Experimental near-field images after Fourier filtering of phase (top) and amplitude (bottom) when resonantly excited and normalized AM power spectra under LCP and RCP light.

For both LCP and RCP excitations, the AM analysis gives a doublet of topological charges. Under LCP the weight is split between $\ell = -2$ and $\ell = 0$, while under RCP the pair moves to $\ell = 0$ and $\ell = +2$. We find good agreement between our experimental results and finite-difference time-domain simulations of the $N = 1$ spiral, analyzed by the same azimuthal Fourier decomposition (Supplementary Note 7). To understand this topological charge splitting of the metasurface, we model the resonant near field of a generalized $N$-armed spiral. For CP illumination, $E_{\mathrm{inc}} \propto \hat{x} + i\sigma\hat{y}$, only the components along the radial direction $\hat{r}(\theta) = \hat{x}\cos\theta + \hat{y}\sin\theta$ excite the out-of-plane mode. Then the projection of the incoming light on the metasurface evaluates to $\hat{r} \cdot E_{\mathrm{inc}} \propto \cos\theta + i\sigma\sin\theta = e^{i\sigma\theta}$, explaining how the resonant near field acquires a geometric phase $\sigma\theta$ as the two bars revolve around the metasurface. The qBIC resonance itself is a stationary solution of the ring lattice, so that its out-of-plane component follows $E_z \propto \cos\varphi$, where $\varphi$ is the local phase of the metasurface. The lattice phase advances by 2π per unit cell in radial direction, and the cosine encodes the sign alternation of the field between neighboring gaps. On an *N*-armed spiral, a full revolution at constant radius crosses *N* unit cells, so that $\varphi = 2\pi r/p + N\theta$. With this, we can model the near field as:

$$E_z \propto cos(\frac{2\pi r}{p} + N\theta) \cdot e^{i\sigma\theta} = \frac{1}{2} e^{i2\pi r/p} e^{i(\sigma+N)\theta} + \frac{1}{2} e^{-i2\pi r/p} e^{i(\sigma-N)\theta}. \quad (3)$$

We note, this is valid where the angle Ψ between the spiral unit cells normal and the radial direction $\Psi = \arctan(Np / 2\pi r)$ is negligibly small, excluding only the most central area of spirals. In this idealized model, the spiral near field decomposes into two helical components of equal weight, one carrying topological charge $\sigma + N$ and the other $\sigma - N$. Their superposition forms a twisted standing wave, similar to how a 1D standing wave is formed by superposition of two counter-propagating plane waves. In our experimental AM spectra, we observe slight deviations in the doublets' peak values, which we attribute to non-ideal illumination conditions induced by the parabolic mirrors used in the s-SNOM measurements.

The phase fronts of the two composing vortices follow the spiral unit-cell arms but travel in opposite radial directions. For a metasurface with an arbitrary number of spiral arms, its topological charges are given by

$$\ell_{\pm} = \pm N + \sigma. \quad (4)$$

This also holds for the concentric ring case ($N = 0$), where both branches coincide at $\ell = \sigma$. The doublet in the spiral metasurfaces' AM spectra is therefore a fingerprint of the standing-wave character of the resonant near fields and allows for an unambiguous assignment of every doublet to the corresponding *N*-armed spiral metasurface and polarization of the mode-exciting light (see Supplementary Note 8). Unlike polaritonic vortices, which are carried by surface waves propagating towards the center of the structure, the near-field textures in our metasurfaces are locally bound to the mode-carrying resonators and therefore form a standing-wave pattern.

To verify our theoretical model and encode higher-order vortices experimentally, which is an essential requirement to be useful for optical communications,[65] we fabricate metasurfaces with $N = 2$, 10, and 20 arms following the same design principles. This demands tailoring of the spiral arms to a radial offset of $N \cdot p$ per revolution, while also applying an angular offset between the unit-cell arms of $\Delta\theta = 2\pi / N$ to maintain the required periodicity of the double-bar unit cells that carry the qBIC resonance field texture. Figure 5a shows the measured topographic image of the spiral metasurface with $N = 2$ arms (left) and its related out-of-plane phase pattern (right). The AM spectrum extracted from the s-SNOM field data (Figure 5b) under RCP illumination shows a doublet at $\ell = -1$ and $\ell = 3$. With increasing numbers of arms, maintaining the periodicity requirement of $p = 4000$ nm between neighboring arms becomes increasingly difficult. At $N = 10$ (Figure 5c, left) the merging of the large number of arms near the center becomes prominent, resulting in an unpatterned Si disk. The corresponding phase profile (Figure 5c, right) partially expands into the unstructured center where a near-field interference pattern is formed. Applying the azimuthal mode decomposition, we obtain the spectrum shown in Figure 5d, which allows for a clear readout of an AM doublet at $\ell = -9$ and $\ell = 11$. Pushing the concept further, at $N = 20$ arms the unstructured central area extends over about 20 µm (Figure 5e, left). In the measured phase map (Figure 5e, right), the spiral pattern extends into the center before the spiral pattern is completely lost. The AM doublet that is revealed in the extracted AM spectrum (Figure 5f) thereby follows the same rule of $\ell_\pm = \pm N + \sigma$, as theoretically predicted.

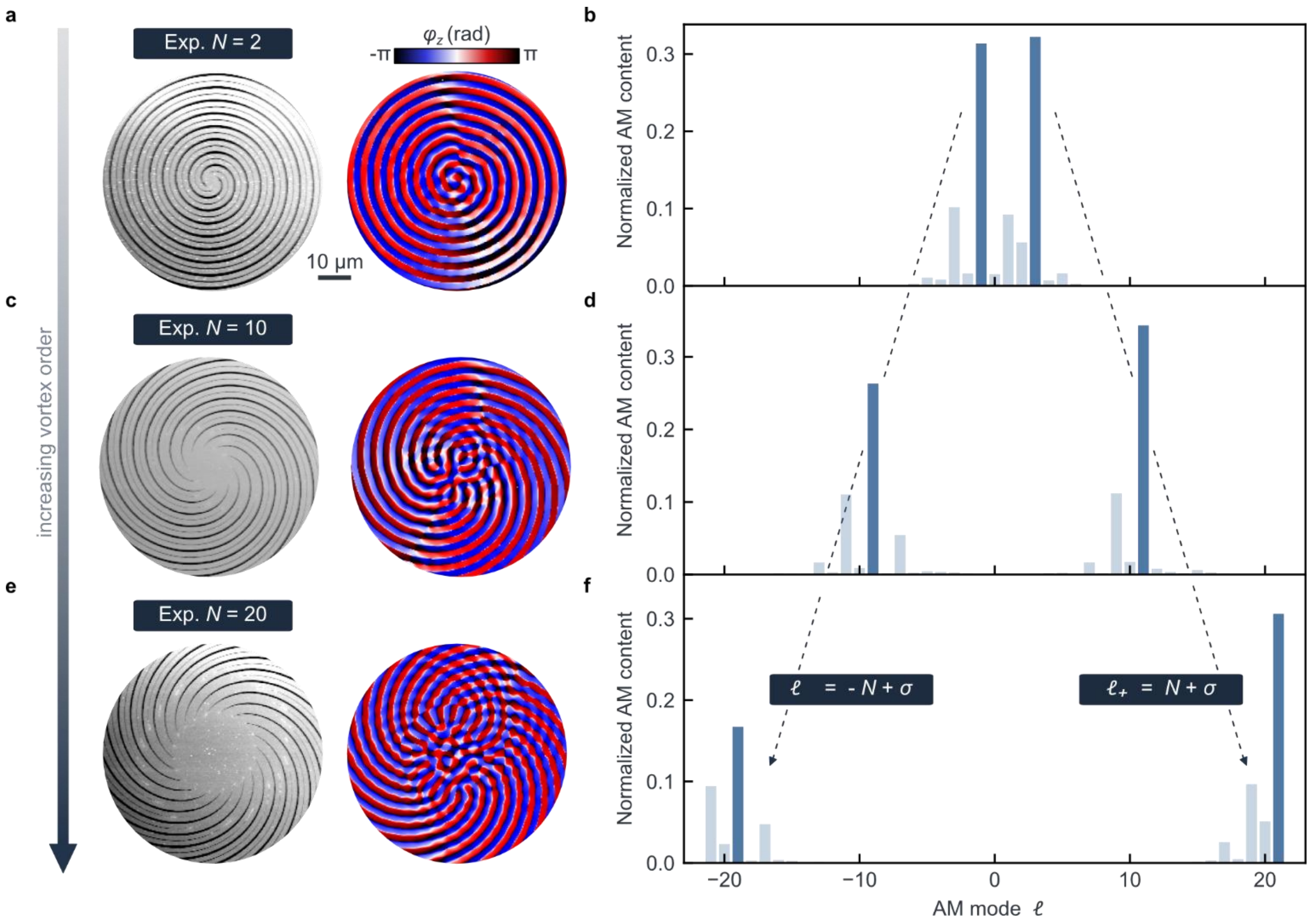

**Figure 5**. Optical vortex control with increasing numbers of arms N and geometrical splitting in AM power spectra under RCP illumination. Experimental topography (left) and phase (right) from s-SNOM measurements of spiral metasurface with (a) N = 2, (c) N = 10 and (e) N = 20, resulting in AM spectra for the respective spiral metasurface showing dominant AM modes at (b) $\ell = -1$ and $\ell = 3$, (d) $\ell = -9$ and $\ell = 11$ and (f) $\ell = -19$ and $\ell = 21$, following the rule $\ell_{\pm} = \pm N + \sigma$.

## CONCLUSION

In conclusion, we have introduced a purely resonance-driven approach to generate near-field topological textures that does not rely on polaritons, but merely on common dielectrics. By bending the unit cells of a 1D displacement grating supporting a qBIC into a ring-shaped structure, the resonance is preserved locally while its field texture is shaped along the layout of the designed non-local metasurface. The generated topology is spectrally locked to the qBIC resonance, whose linewidth can be independently controlled by the asymmetry parameter $\alpha$. Phase-resolved s-SNOM measurements of our Si ring metasurfaces reveal optical near-field vortices in the structures. Expanding these design principles to Archimedean spirals, our platform enables the generation of AM doublets at $\ell_{\pm} = \pm N + \sigma$, realized with up to $N = 20$ arms. The derived standing-wave model predicts the dominant peak positions in the measured AM spectra through the superposition of two helical components, distinguishing our resonant textures from their polaritonic counterparts with propagating nature.

Our approach paves the way for the realization of non-local and stationary topological phenomena, such as the Néel-type target skyrmion demonstrated numerically here, which are generated coherently and locally rather than requiring the phase-engineered launching mechanisms used for their polaritonic counterparts. Since our topological textures are generated by a photonic resonance rather than a material-dispersion induced hybrid excitation, the operating spectral window is not restricted to the polariton dispersion, but can be set using the geometric scaling factor $S$. The concept is thereby transferable to other spectral ranges with suitable low-loss dielectric materials. Especially towards the visible and UV, phonon polaritons are unavailable and plasmonic metals such as gold become increasingly lossy.[42] Our approach, in contrast, relies purely on high-index, low-loss dielectrics, making materials such as silicon nitride, hafnium oxide or zirconium oxide viable options. These offer wide bandgaps and therefore extend the application range into the UV.[66] The radial periodicity of the metasurface shrinks from $p = 4000$ nm using silicon in the mid-IR to $p = 229$ nm using silicon nitride to shift the out-of-plane mode to a wavelength of 298 nm (33542 $cm^{-1}$) (Supplementary Note 9), enabling the compression of the topology with its full information content and maximizing on-chip information density. We further envision the extension of our approach to the far field by introducing a suitable coupling mechanism such as an additional long-period modulation reminiscent of diffractive non-local qBIC metasurfaces.[67] This opens pathways to topologically structured and coherent thermal emission, since patterned qBIC metasurfaces have been shown to imprint a geometric phase on incoherent emission from a heated surface.[68] Overall, the geometric control over near-field topologies opens a route to near-field information encoding and on-chip topological photonics, and could potentially serve as versatile and scalable source of coherent, topologically structured thermal radiation.

## METHODS

**Numerical Modeling**. Simulations were conducted using a commercial finite-difference time-domain software (Ansys Lumerical). The boundary conditions were set to periodic for the 1D and sinusoidal gratings, whereas for the ring and spiral designs the entire metasurfaces were simulated with perfectly matched layers due to the absence of a grid periodicity. The latter were simulated with 30 unit cells to ensure mode formation and to minimize edge effects. Unless otherwise stated, all structures were excited with CP light. Radial polarization was used for Figure 1c and Supplementary Note 3, and linear polarization for Supplementary Note 1. The refractive indices of the silicon resonators and the substrate were set to $n_{res} = 3.4$ and $n_{sub} = 1.33$, respectively. For Supplementary Note 9, the refractive index of silicon nitride was retrieved from Beliaev *et al.*[69] Far-field spectra were recorded by a local transmittance monitor spanning 20 · 20 µm and positioned 100 µm laterally from the metasurface center, mimicking the spatially resolved readout of the experimental hyperspectral imaging microscope (see Optical Far-Field Measurements). The numerically calculated near fields were extracted in a plane 5 nm above the surface of the resonators for comparison with the experimental s-SNOM measurements (see Optical Near-Field Measurements).

**Fabrication.** The qBIC metasurfaces were fabricated through a previously developed inverse multistep fabrication procedure[70]: A 1230 nm thick amorphous Si film was deposited onto a $CaF_2$ substrate by plasma-enhanced chemical vapor deposition using a PlasmaPro 100 reactor (Oxford Instruments), followed by depositing a 110 nm chromium hard mask using an Amod physical vapor deposition system (Angstrom Engineering). After the application of an adhesion promoter (AR 300-80 new, Allresist), a positive electron-beam resist (AR-P 6200.13, Allresist) was spin-coated at 3000 rpm and baked at 175 °C for 5 minutes, and topped with a conductive polymer (Espacer 300Z, Showa Denko K.K.). Subsequently, the patterns were inversely written by electron beam lithography (eline Plus system, Raith) operating at 20 kV with a 15 µm aperture and the development was conducted for 60 s in amyl acetate and 15 s in a MIBK:IPA 1:9 solution. The resulting polymer patterns were then translated from the resist into the chromium hard mask and subsequently into the Si layer via reactive ion etching (PlasmaPro 100, Oxford Instruments), before the remaining chromium was eventually removed, yielding the final devices.

**SEM Imaging.** After all optical measurements were performed, SEM images were taken with a Gemini device (Zeiss) operating at 2 kV with a 30 µm aperture. Beforehand, a few nm thin palladium-gold film was sputtered on the samples to reduce charging effects and to increase the contrast.

**Optical Far-Field Measurements.** Spatially resolved far-field spectra were acquired using a hyperspectral IR imaging microscope (Spero-QT, Daylight Solutions) equipped with three tunable quantum cascade lasers operated in transmittance mode. The sample was illuminated from 952 $cm^{-1}$ to 1800 $cm^{-1}$ in 2 $cm^{-1}$ steps through a 4× magnification objective (NA = 0.15), providing a 2 · 2 mm field of view with a resolution of 480 · 480 pixels and a pixel size of ≈ 4 · 4 µm and the spectra were normalized to a bare $CaF_2$ substrate. A quarter-wave plate was used to generate LCP and RCP light at the resonance frequency.

**Optical Near-Field Measurements.** All near-field experiments were conducted with a commercial s-SNOM system (neaSCOPE, Attocube Systems). The laser source was a continuous-wave quantum cascade laser (MirCat, Daylight Solutions), tunable from 1315 $cm^{-1}$ to 1755 $cm^{-1}$, which covers the entire range of interest. All measurements were done in transmission mode in order to ensure normal incidence to properly excite the qBIC mode and the light passed through a linear polarizer followed by a quarter-wave plate to obtain either LCP or RCP light. To study the topological properties of qBIC-driven optical vortices, both near-field scattering amplitude $|E_z|$ and phase $\varphi_z$ need to be obtained, which requires the use of pseudo-heterodyne (PsHet) detection.[63] To accomplish this, the beam is first split in two by passing through a beam splitter. One part is loosely focused onto the metasurface and the probing AFM tip at 0° incidence from below (through the $CaF_2$ substrate) via a parabolic mirror. The tip is operated in tapping mode, oscillating at a frequency of around 250 kHz. For all experiments, we used metal-coated (platinum/iridium) AFM tips (Arrow-NCPt, NanoWorld). The light backscattered from the tip is then re-collected through a second parabolic mirror. The second part of the beam passes through a delay stage, in which two mirrors vibrate with a defined mirror amplitude to achieve decoupling of optical amplitude and phase. When conducting PsHet interferometry,[71] the frequency of vibration for these two mirrors (around 300 Hz) is much lower than the tip frequency. Both beams are recombined at a second beam splitter and measured with a liquid nitrogen-cooled mercury cadmium telluride detector. To distinguish the near fields from far-field background scattering, the signal is demodulated at higher tip harmonics ($n > 2$ for all measurements, using $n = 3$ in analysis), ensuring that only the local near fields generated on top of each resonator are detected. In addition, PsHet enables the elimination of the multiplicative background contribution otherwise present in all s-SNOM experiments. The reference mirror introduces side peaks next to the main tip harmonics, which do not contain the multiplicative background, thus demodulation at the sideband frequencies of sufficiently high $n$ allows for conducting near-field measurements without any far-field contributions.[63]

**Supporting Information Available**

Supplementary Notes 1–9: qBIC near fields under linear polarization, *Q*-factor scaling, simulated near-field target skyrmion, Fourier filtering, azimuthal Fourier decomposition, symmetric-metasurface control, simulated spiral-mode analysis, standing-wave nature of Archimedean spiral metasurface, silicon nitride metasurface for the UV.

**Corresponding Authors**

**Andreas Tittl** - Institute of Photonics, Hamburg University of Technology, 21073 Hamburg, Germany; Chair in Hybrid Nanosystems, Nanoinstitute Munich, Department of Physics, Ludwig-Maximilians-Universität, 80539 Munich, Germany. Email: andreas.tittl@tuhh.de
https://orcid.org/0000-0003-3191-7164

**Enrico Baù** - Chair in Hybrid Nanosystems, Nanoinstitute Munich, Department of Physics, Ludwig-Maximilians-Universität, 80539 Munich, Germany. Email: enrico.bau@physik.uni-muenchen.de
https://orcid.org/0009-0000-6251-1875

**Authors**

**Jonas Biechteler** - Institute of Photonics, Hamburg University of Technology, 21073 Hamburg, Germany; Chair in Hybrid Nanosystems, Nanoinstitute Munich, Department of Physics, Ludwig-Maximilians-Universität, 80539 Munich, Germany.
https://orcid.org/0009-0009-1975-4420

**Michael Hirler** - Institute of Photonics, Hamburg University of Technology, 21073 Hamburg, Germany; Chair in Hybrid Nanosystems, Nanoinstitute Munich, Department of Physics, Ludwig-Maximilians-Universität, 80539 Munich, Germany.
https://orcid.org/0009-0008-4154-5791

**Connor Heimig** - Institute of Photonics, Hamburg University of Technology, 21073 Hamburg, Germany; Chair in Hybrid Nanosystems, Nanoinstitute Munich, Department of Physics, Ludwig-Maximilians-Universität, 80539 Munich, Germany.
https://orcid.org/0009-0001-6820-2157

**Martin Barkey** - Institute of Photonics, Hamburg University of Technology, 21073 Hamburg, Germany; Chair in Hybrid Nanosystems, Nanoinstitute Munich, Department of Physics, Ludwig-Maximilians-Universität, 80539 Munich, Germany.
https://orcid.org/0000-0002-2399-0698

**Haoran Ren** - School of Physics and Astronomy, Monash University, Clayton, Victoria 3800, Australia.
https://orcid.org/0000-0002-2885-875X

**Stefan A. Maier** - School of Physics and Astronomy, Monash University, Clayton, Victoria 3800, Australia; Department of Physics, Imperial College London, London SW7 2AZ, United Kingdom.
https://orcid.org/0000-0001-9704-7902

**Author Contributions**

*J.B. and M.H. contributed equally to this work. J.B., M.H. and E.B. developed the idea. J.B., M.H. and C.H. performed the simulations. J.B. fabricated the samples. J.B., E.B. and M.B. conducted the optical characterization, while M.H. performed SEM imaging. J.B., M.H. and E.B. did the data analyses. H.R., S.A.M. and A.T. supervised the project. All authors reviewed and approved the final manuscript.

**Conflict of Interest**

The authors declare no competing interest.

**Data Availability Statement**

The data that support the findings of this study are available from the corresponding authors upon reasonable request.

**Acknowledgment**

Funded by the European Union (EIC, OMICSENS, 101129734, EIC, NEHO, 101046329, ERC, METANEXT, 101078018). Views and opinions expressed are however those of the author(s) only and do not necessarily reflect those of the European Union or the European Research Council Executive Agency. Neither the European Union nor the granting authority can be held responsible for them. This project was also funded by the Deutsche Forschungsgemeinschaft (DFG, German Research Foundation) under Germany's Excellence Strategy (EXC 3120/1 – 533771286 and EXC 2089/1–390776260) and the Emmy Noether

Programme (TI 1063/1), the Bavarian program Solar Energies Go Hybrid (SolTech) and the Center for NanoScience (CeNS). H.R. acknowledges funding support from the Australian Research Council Grant (FT250100565). S.A.M. acknowledges the Australian Research Council (DP250102064) and the Lee-Lucas Chair in Physics.

# Supporting Information for

# Shaping Near Fields of Photonic Modes into Topological Textures

*Jonas Biechteler*[*,1,2], *Michael Hirler*[*,1,2], *Connor Heimig*[1,2]*, Martin Barkey*[1,2]*, Haoran Ren*[3]*,*

*Stefan A. Maier*[3,4], *Enrico Baù*[‡,2] *and Andreas Tittl*[†,1,2]

[1]Institute of Photonics, Hamburg University of Technology, 21073 Hamburg, Germany.

[2]Chair in Hybrid Nanosystems, Nanoinstitute Munich, Department of Physics, Ludwig-Maximilians-Universität, 80539 Munich, Germany.

[3]School of Physics and Astronomy, Monash University, Clayton, Victoria 3800, Australia.

[4]Department of Physics, Imperial College London, London SW7 2AZ, United Kingdom.

*These authors contributed equally to this work.

[†]Email: andreas.tittl@tuhh.de

[‡]Email: enrico.bau@physik.uni-muenchen.de

## Supplementary Note 1 - Electric near-fields under linear polarization

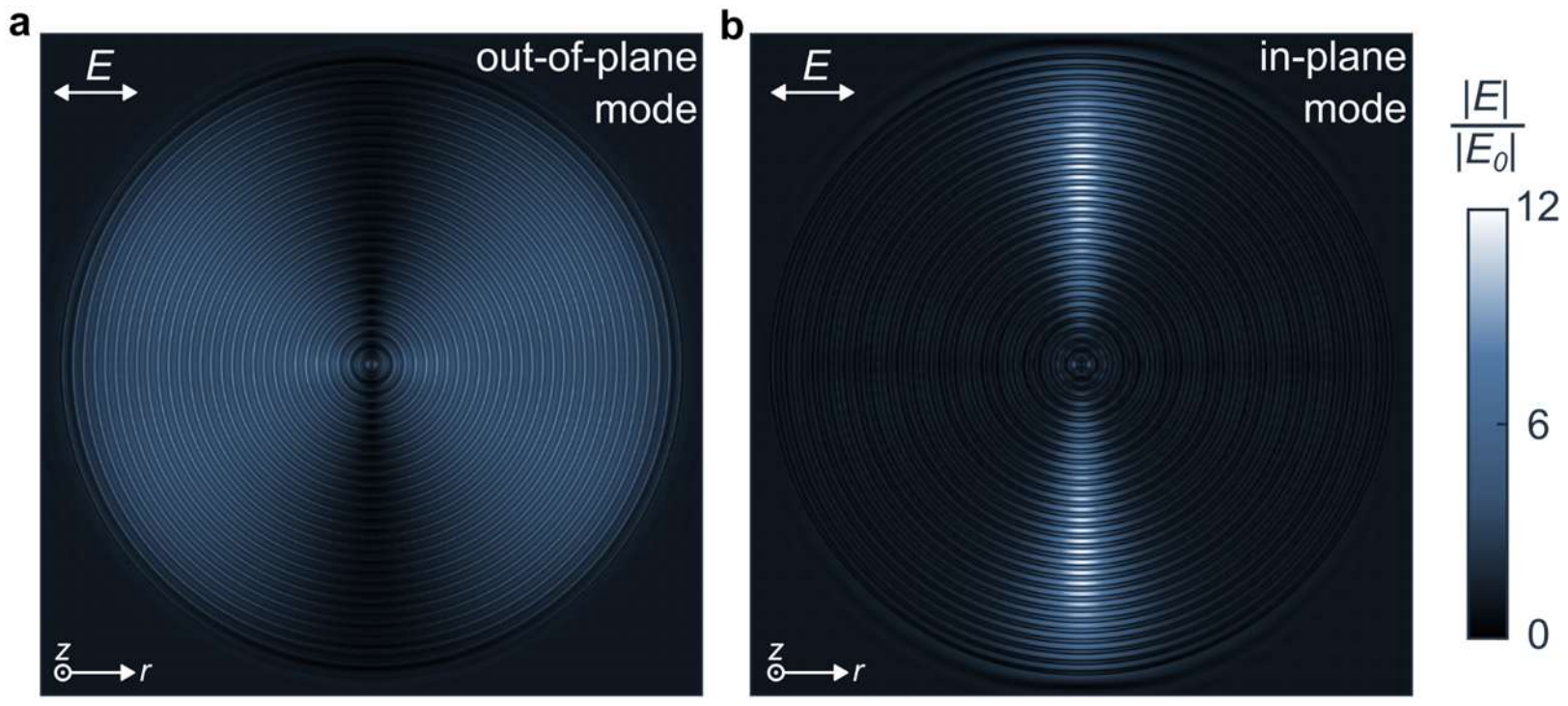

**Figure S1**. Electric near fields of a concentric ring metasurface. Top view of the concentric ring metasurface shows the simulated electric near-field distribution for $\alpha = 0.4$ under linearly polarized light for the (a) out-of-plane and (b) in-plane modes, evaluated 5 nm above the resonators. The out-of-plane mode is excited predominantly where the polarization is oriented across the bars, whereas the in-plane mode is excited predominantly where it is oriented along them.

## Supplementary Note 2 - *Q* factor tuning via asymmetry parameter

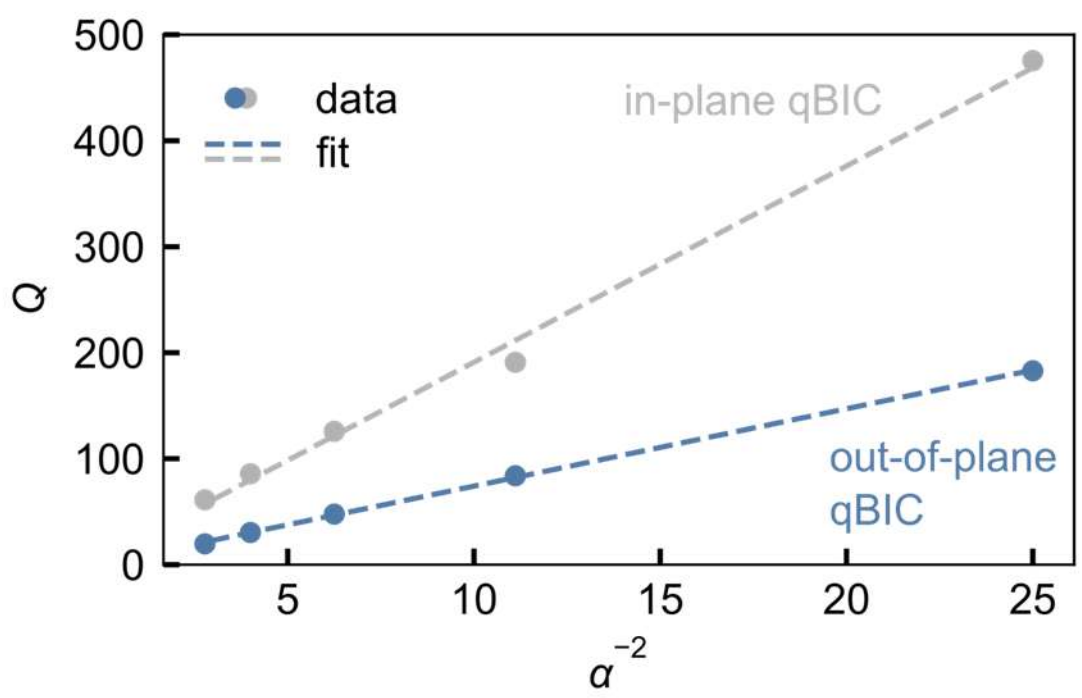

**Figure S2**. Dependence of the *Q* factor on the asymmetry parameter $\alpha$. In the absence of intrinsic losses, the *Q* factors of the out-of-plane and in-plane qBICs follow the well known relationship $Q \propto \alpha^{-2}$.[1] The *Q* factor values were extracted from simulations by fitting a Fano line shape to the transmittance spectra. The linear fits follow $7.30\alpha^{-2} + 1.15$ for the out-of-plane and $18.51\alpha^{-2} + 5.90$ for the in-plane mode, respectively.

## Supplementary Note 3 - Target-skyrmion case

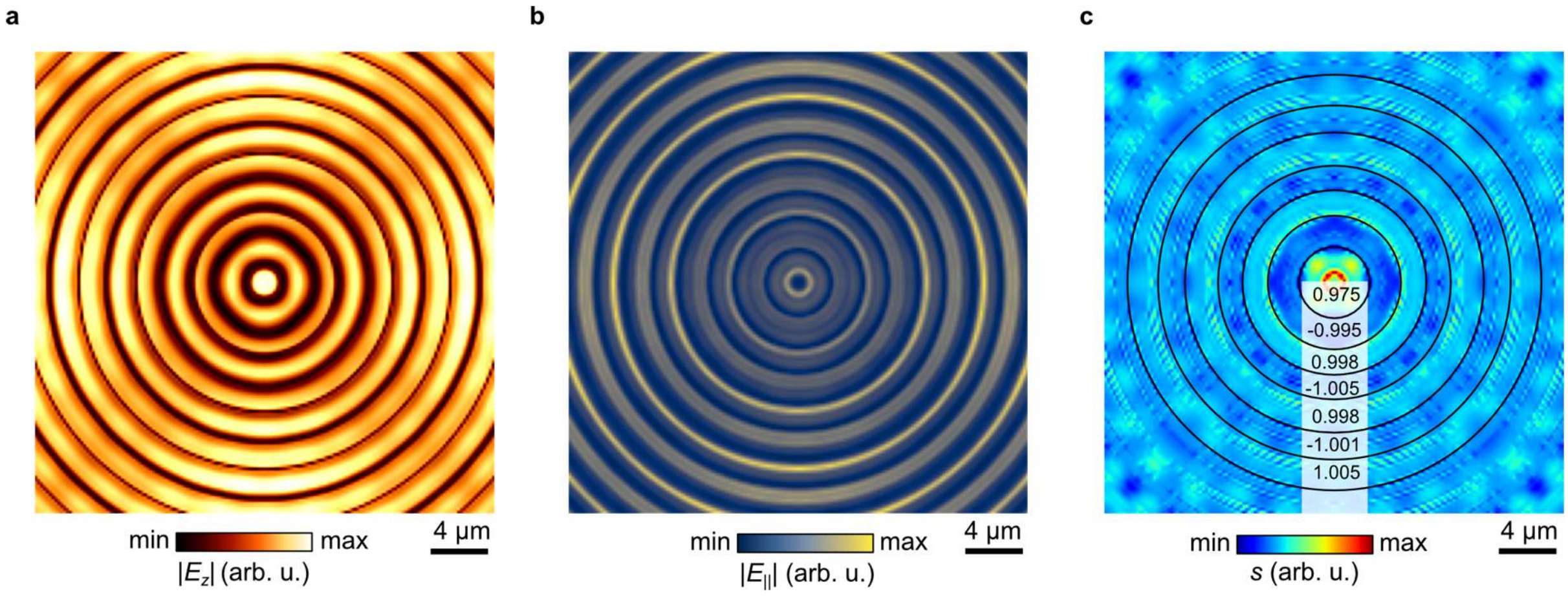

**Figure S3**. Near-field target skyrmion in a concentric ring metasurface under radially polarized excitation. (a) Simulated out-of-plane field amplitude $|E_z|$ and (b) in-plane field amplitude $|E_{||}|$, both evaluated 5 nm above the resonators. (c) Skyrmion number density $s$. The numbers denote the calculated skyrmion number $S_T$ for each area obtained by integrating $s$ over the successive rings of width $p/2 \pm (g_2\text{-}g_1)/2$ delimited by the black contours, alternating between numbers very close to the ideal scores of +1 and −1. This sequence of alternating shells constitutes a Néel-type target skyrmion.[2] For an animated version, see Supplementary Movie M1.

## Supplementary Note 4 - Fourier filtering of near-field data

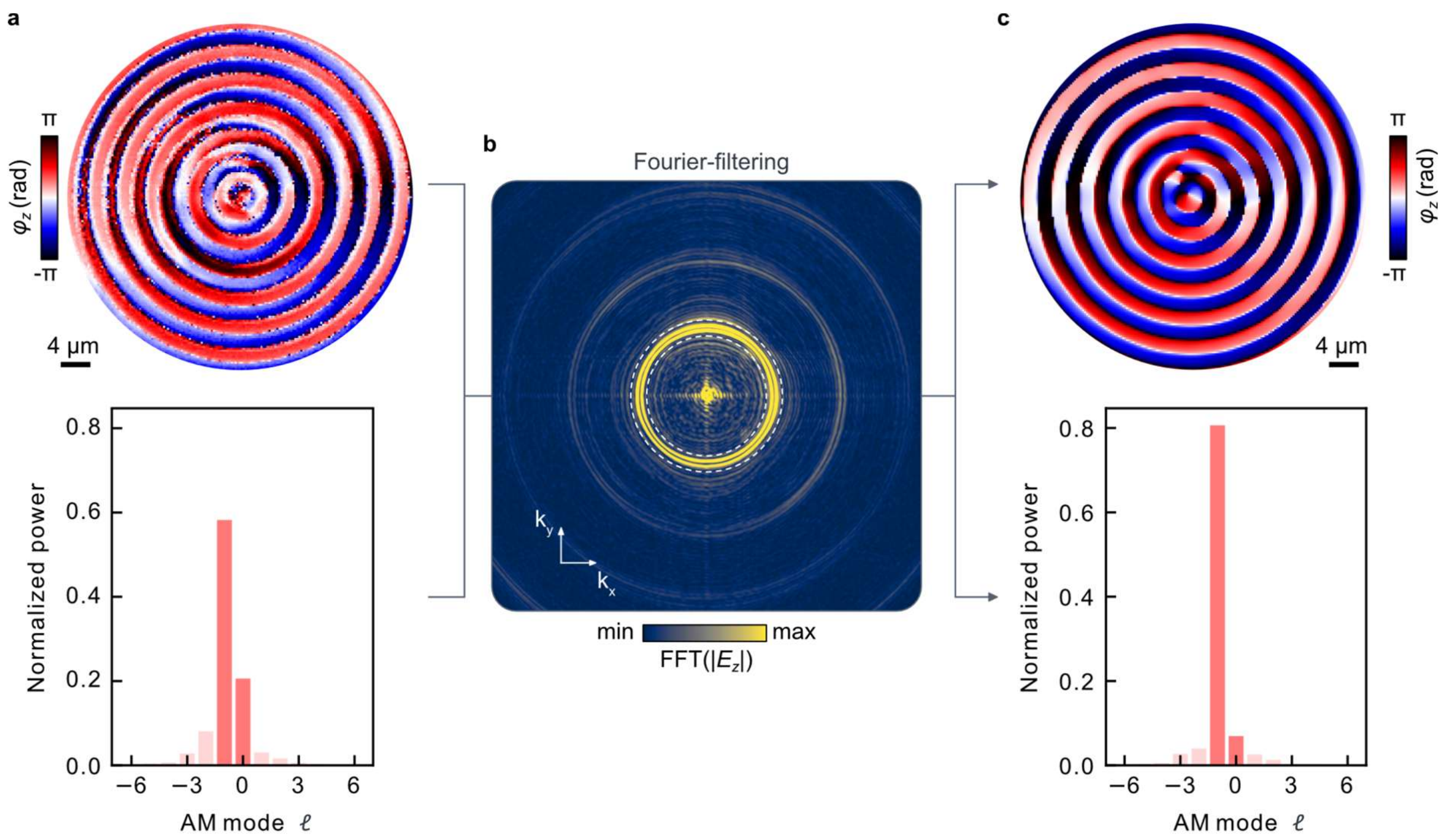

**Figure S4**. Fourier-filtering of the measured near-field data. (a) Phase map of raw near-field data measured using s-SNOM (top) and respective AM spectrum (bottom) for the concentric ring metasurface under LCP excitation. (b) The data are then Fourier-transformed and filtered accordingly to macht the unit-cell periodicity in radial direction indicated by the dashed lines. (c) After inverse Fourier-transformation, the recovered phase map (top) shows improved visual clarity, removing measurement induced artifacts, as well as background scattering, and increased AM spectrum purity (bottom).

## Supplementary Note 5 - Azimuthal Fourier decomposition

The AM spectra shown in the main text (Figures 3, 4 and 5) are obtained by an azimuthal Fourier decomposition of the near fields. For this, the complex out-of-plane near field $E_z = |E_z|\, e^{i\varphi z}$, which was measured by s-SNOM, is expanded into a complete basis of helical modes $e^{i\ell\theta}$ on rings centered around the phase singularity[3]:

$$E_z(r,\theta) = \sum_\ell c_\ell(r)\, e^{i\ell\theta} \,. \tag{S1}$$

The expansion coefficients $c_\ell(r)$ of each topological charge $\ell$ are calculated as the angular Fourier transform of the field on each circle of radius $r$

$$c_\ell(r) \; = \frac{1}{2\pi} \int_0^{2\pi} E_z(r,\theta)\, e^{-i\ell\theta} d\theta \; . \tag{S2}$$

The normalized weight of the $\ell$-th helical component is then obtained by integrating the squared modulus of the coefficients over the radial coordinate

$$P_\ell = \frac{1}{W} \int_0^R |c_\ell(r)|^2 \, r \, dr \; , \tag{S3}$$

using the integrated squared out-of-plane field amplitude for normalization

$$W = \sum_\ell \int_0^R |c_\ell(r)|^2 \, r \, dr = \frac{1}{2\pi} \int_0^R \int_0^{2\pi} |E_z(r,\theta)|^2 r \, d\theta \;\; dr \;\; . \tag{S4}$$

The maximum integration radius $R \approx 45$ µm is set by the s-SNOM scan area.

## Supplementary Note 6 - Symmetric metasurface control

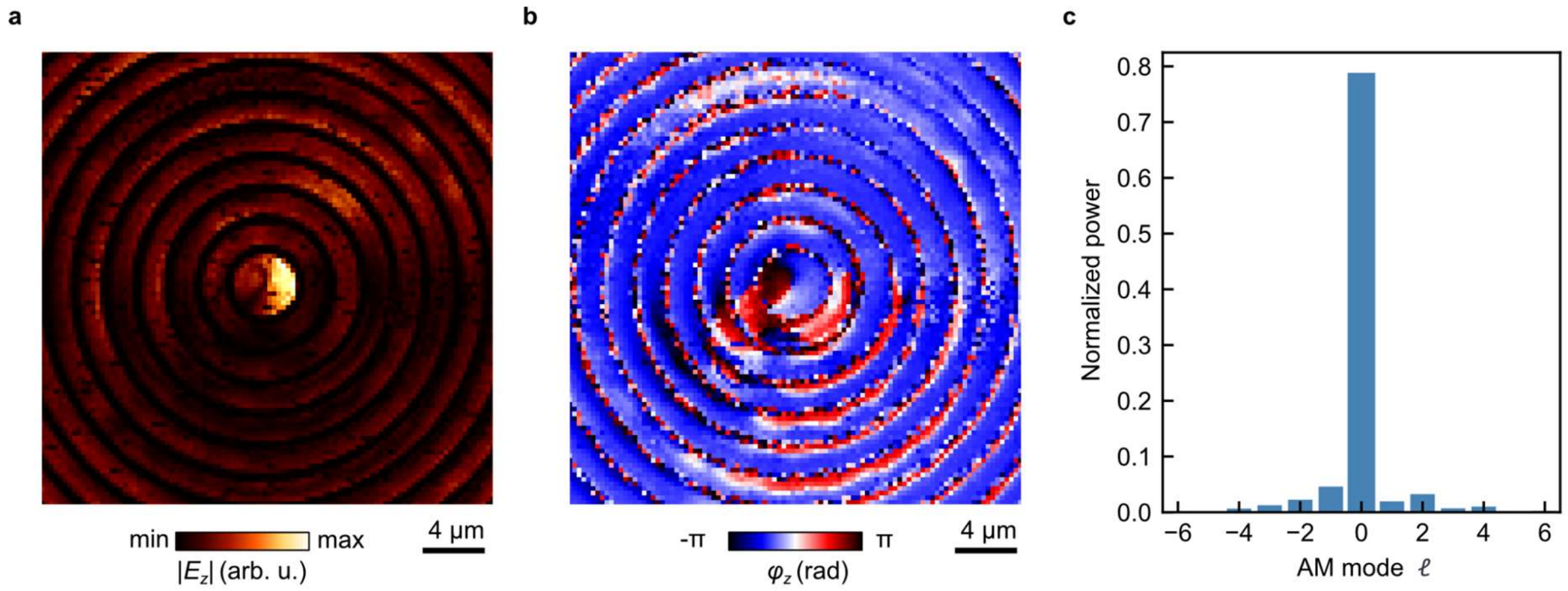

**Figure S5**. Absence of the near-field vortex in the symmetric ($\alpha = 0$) metasurface. (a) Measured out-of-plane near-field amplitude $|E_z|$ and (b) phase $\varphi_z$ of the concentric ring metasurface under LCP illumination at 1622 $cm^{-1}$. (c) Corresponding AM power spectrum, dominated by $\ell = 0$.

## Supplementary Note 7 - Simulation of Archimedean spiral metasurfaces

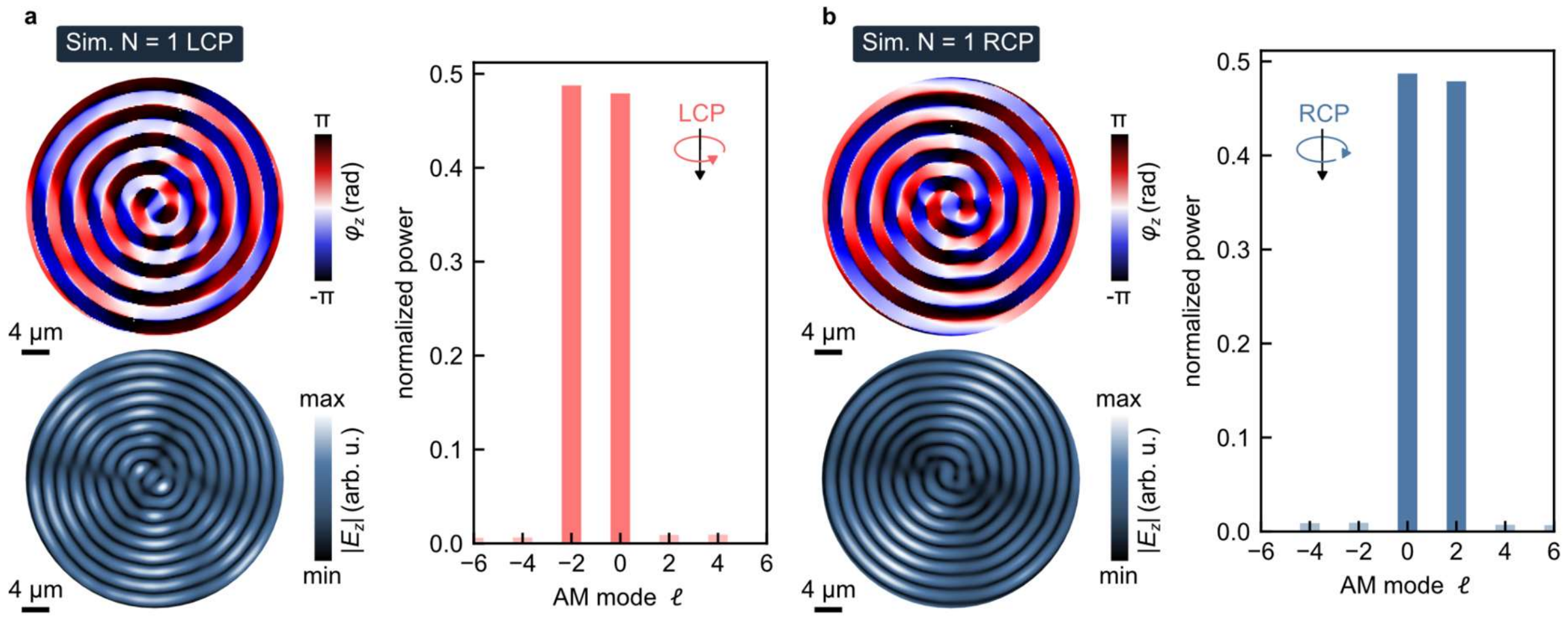

**Figure S6**. Simulations of Archimedian spiral metasurface. (a) Simulated phase (top) and amplitude (bottom) of the near-fields of a $N = 1$ spiral metasurface under LCP excitation with AM at $\ell = 0, -2$. (b) Corresponding RCP result, with peaks at $\ell = 0, +2$.

## Supplementary Note 8 - Standing-wave nature of Archimedean spiral metasurfaces

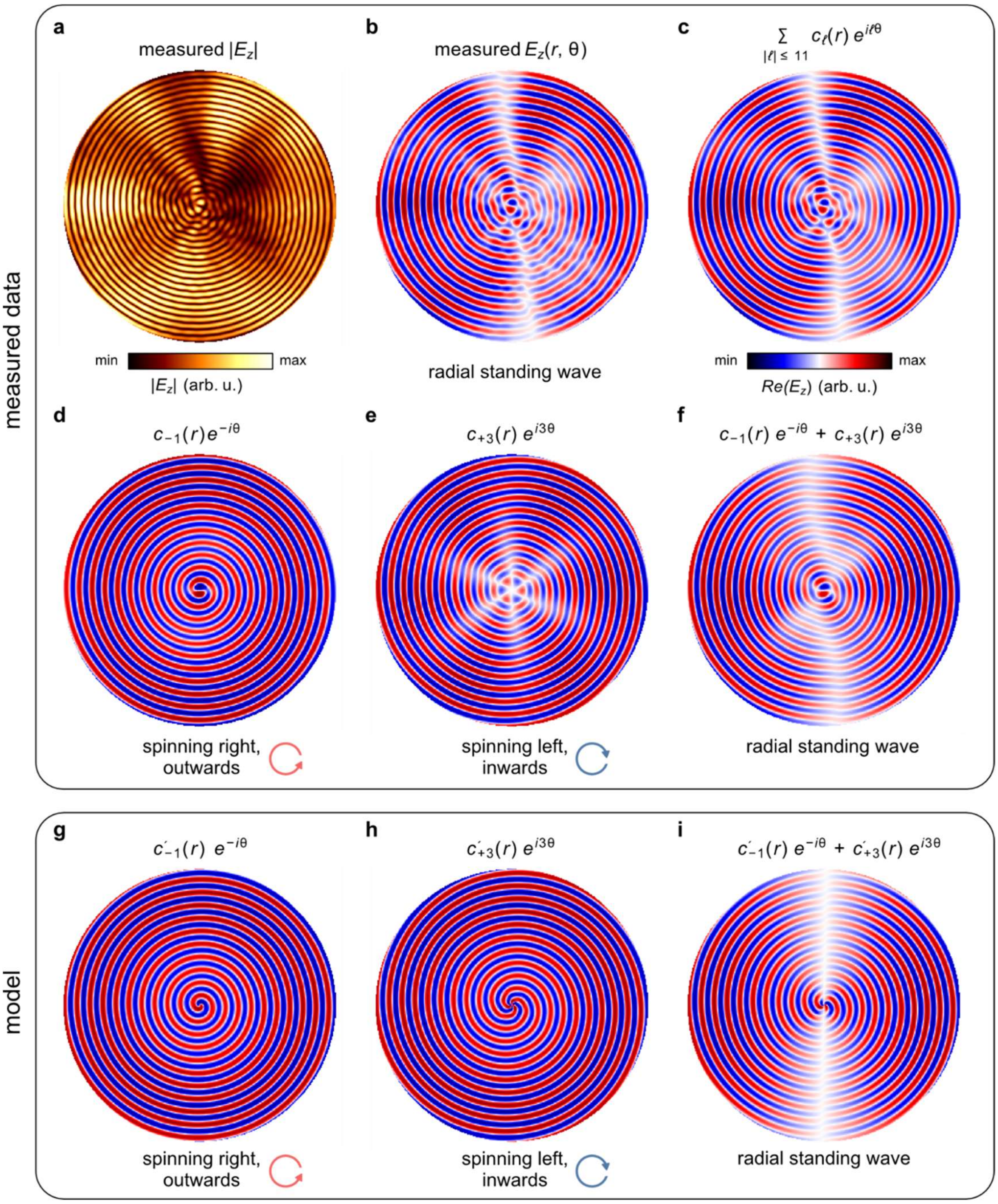

**Figure S7.** Helical-mode decomposition of the measured out-of-plane near field of the $N = 2$ spiral metasurface, compared with the standing-wave model. The measured field is expanded as $E_z(r, \theta) = \sum_\ell c_\ell(r)\, e^{i\ell\theta}$, with $c_\ell(r) = (1/2\pi)\int E_z(r, \theta)\, e^{-i\ell\theta}\, d\theta$ the angular Fourier transform of the field on the circle of radius $r$. Panels (b-i) show the real part of the corresponding term, $\mathrm{Re}[c_\ell(r)e^{i\ell\theta}]$. (a) Measured amplitude $|E_z|$. (b) Real part of the measured field. (c) Sum over $c_\ell(r)$ for $|\ell| \leq 11$, reproducing (b). (d,e) The two strongest AM components of the measurement, $c_{-1}(r)\, e^{-i\theta}$ and $c_{+3}(r)\, e^{i3\theta}$. (f) Their superposition. (g-i) Helical modes $c'_\ell(r)\, e^{i\ell\theta}$ for $\ell = -1$, 3 and their sum obtained using the theoretical model $E_z = [a\cdot e^{i\varphi} + b\cdot e^{-i\varphi}]\, e^{i\sigma\theta}$, with $\varphi = 2\pi r/p + N\theta$; fitting constants $a, b = (1 \pm \beta)/2$, $N = 2$, $\sigma = +1$, $\beta = 0.02$ and the radial periodicity of the model $p_r = 4082$ nm, which was fitted to the measured radial amplitude profile. For an animated version, see Supplementary Movie M2.

**Supplementary Note 9 - Silicon nitride concentric ring metasurface for the UV**

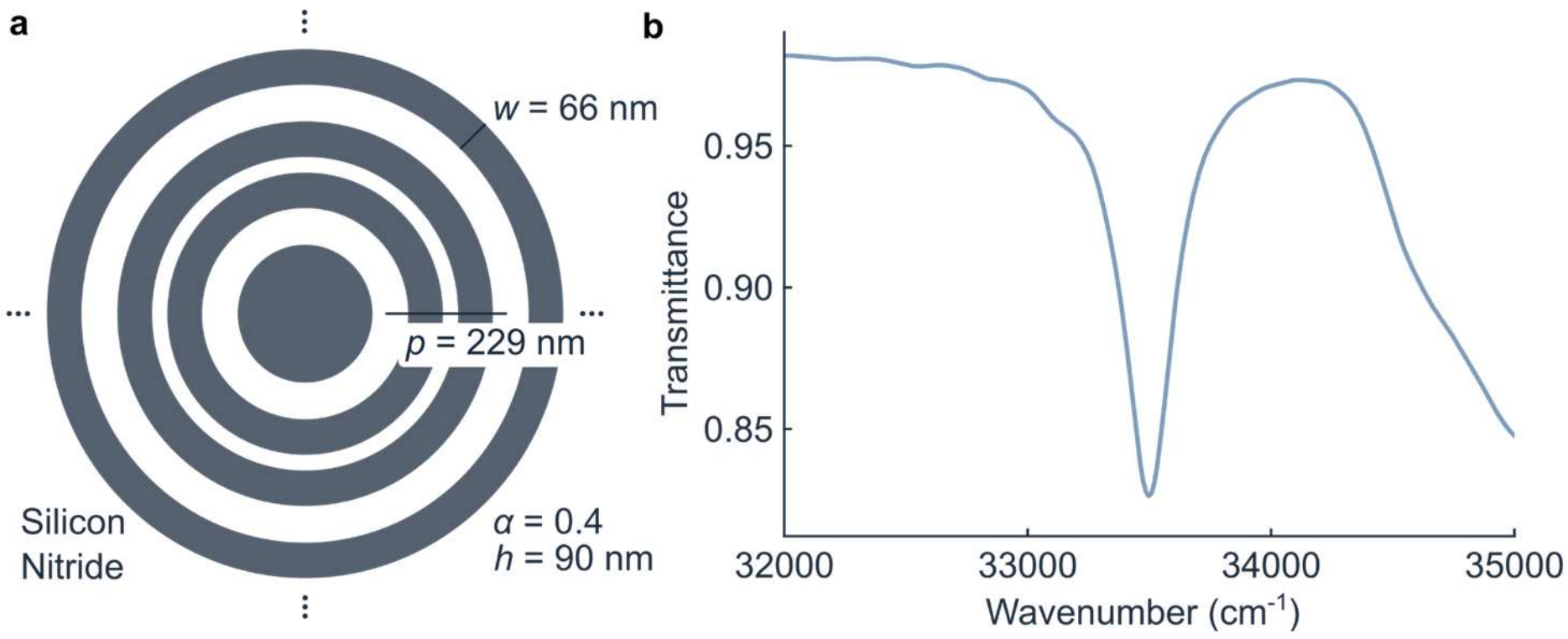

**Figure S8.** Silicon nitride concentric ring metasurface for the UV. (a) Schematic of the concentric ring metasurface (first four rings). (b) Silicon nitride allows to scale the out-of-plane qBIC to UV frequencies (here 33542 $cm^{-1}$) for maximizing information density. The refractive index was taken from Beliaev *et al.*[4]